\documentstyle[a4,12pt,epsfig]{article}
\def\be{\begin{equation}}
\def\ee{\end{equation}}
\def\bq{\begin{eqnarray}}
\def\eq{\end{eqnarray}}
\newcounter{saveeqn}
\newcounter{App} 
\newcommand{\app}{%
\stepcounter{App}%
\setcounter{saveeqn}{\value{equation}}%
\setcounter{equation}{0}%
\renewcommand{\theequation}{\Alph{App}\arabic{equation}} }
\newcommand{\appende}{%
\setcounter{equation}{\value{saveeqn}}%
\renewcommand{\theequation}{\arabic{equation}}  }
\begin{document}
\thispagestyle{empty}
\setcounter{page}{0}
\begin{flushright}
CERN-TH/98-58\\
WUE-ITP-98-006\\
\end{flushright}
\vspace*{\fill}
\begin{center}
{\Large\bf The Scalar $B\to \pi$ and $D \to \pi$ Form Factors in QCD}\\
\vspace{2em}
{\bf A. Khodjamirian$^{a,1}$, R. R\"uckl$^{a,b}$, C.W. Winhart$^a$}\\
\vspace{2em}
{$^a$ \small Institut f\"ur Theoretische Physik,
Universit\"at W\"urzburg, D-97074 W\"urzburg, Germany}\\
{$^b$ \small Theory Division, CERN, CH-1211 Gen\`eve 23, Switzerland}\\
\end{center}
\vspace*{\fill}

\begin{abstract}

QCD sum rules on the light-cone are derived for the sum 
$f^+ + f^-$ of the $B\rightarrow \pi$ and 
$D\rightarrow \pi$ form factors 
taking into account contributions up 
to twist four. Combining the results with the corresponding
$f^+$ form factors calculated previously by the same method, we obtain
the scalar form factors $f^0$. Our sum rule predictions
are compared with lattice results, current-algebra 
constraints, and quark-model calculations. 
Furthermore, we calculate  
decay distributions and the integrated width for the semileptonic
decay  $B \rightarrow \pi \bar{\tau}\nu_\tau$  which is sensitive to  
$f^0$. Finally, the dependence of the sum rules on the heavy quark
mass and the asymptotic scaling laws are discussed.

\end{abstract}
 
\vspace*{\fill}
 
\begin{flushleft}
$^1$ {\small \it On leave from 
Yerevan Physics Institute, 375036 Yerevan, Armenia } \\
\end{flushleft}
 
\newpage

\section{Introduction}
The weak transition $B\rightarrow \pi$ plays an exceptional role
in $B$ physics, in particular at future $B$ factories.
The amplitude of this transition  
is given by the hadronic matrix element 
\be
\langle\pi(q)\mid \bar{u}\gamma_\mu b\mid B(p+q)\rangle
=2f^+(p^2) q_\mu +(f^+(p^2)+f^-(p^2)) p_\mu ~,
\label{def}
\ee
where $p+q$ and $q$ denote the initial and final state four-momenta,
respectively, $\bar{u}\gamma_\mu b$ is the relevant weak vector
current, and $f^{\pm}$ are the two independent form factors.
The form factor $f^+(p^2)$ was calculated
in \cite{BKR,BBKR} using the technique of QCD 
sum rules on the light-cone.
Recently, one has also computed the
perturbative QCD corrections to $f^+$ \cite{KRWY,Bagan}.  

In the present paper 
we complete the calculation 
of the matrix element (\ref{def}) 
by deriving the corresponding 
light-cone sum rule for the sum of form factors 
$f^+ +f^-$. This quantity turns out to be  
a pure higher-twist effect. The leading twist-2 contribution vanishes 
kinematically. From $f^+ +f^-$ 
and $f^+$, one can construct
the scalar form factor 
\be
f^0(p^2) =\left(1-\frac{p^2}{m_B^2-m_\pi^2}\right) f^+(p^2)
+\frac{p^2}{m_B^2-m_\pi^2}\left(f^+(p^2) + f^-(p^2)\right) ~, 
\label{f0}
\ee
which determines the matrix element 
of the divergence of the weak vector current:
\be
\langle\pi(q)\mid \partial^\mu (\bar{u}\gamma_\mu b)\mid B(p+q)\rangle
=(m_B^2-m_\pi^2)f^0(p^2)~. 
\label{def1}
\ee
Using the results on $f^+$ and $f^0$  we predict
the momentum transfer and lepton energy 
distributions as well as the width of the semileptonic decay 
$B \to \pi \bar{\tau} \nu_\tau$. 
As a by-product, we also obtain
the analogous form factors of the $D\rightarrow \pi$ transition.
 
Furthermore, we investigate 
the heavy-mass dependence of heavy-to-light form factors. 
The asymptotic scaling laws are 
determined and found to differ at small and 
large momentum transfer.
The origin of this difference is explained in detail. We also 
study  the approach to the heavy-quark limit numerically
and show that it is reached very slowly. 
Moreover, the behaviour beyound the physical $b$-quark mass 
turns out to be very  sensitive to the scale
dependence of the pion wave functions. 

The paper is organized as follows.
In sect. 2, we derive the light-cone sum rule for  $f^+ + f^-$ and compare
it with the corresponding sum rule for $f^+$. 
The numerical analysis of the new sum rule and the resulting prediction of 
the scalar form factor $f^0$ is presented in sect. 3.
Sect. 4 is devoted to the semileptonic decay  
$B \rightarrow \pi \bar{\tau} \nu_\tau$,
and sect. 5 to the heavy-mass dependence of the form factors. 
Our conclusions are summarized in sect. 6.

\section{Light-cone sum rule for $f^+ + f^-$}

In order to obtain the QCD sum rule for the form factor combination
$f^+ + f^-$ appearing in (\ref{def}),
we follow the method applied to $f^+$ and explained in detail 
in \cite{BKR,BBKR}. The main object of investigation is
the vacuum-pion correlation function 
\bq
F_\mu (p,q)&=&
i \int d^4x~e^{ipx}\langle \pi(q)\mid T\{\bar{u}(x)\gamma_\mu b(x),
\bar{b}(0)i\gamma_5 d(0)\}\mid 0\rangle \nonumber\\ 
&=&F(p^2,(p+q)^2)q_\mu +\widetilde{F}(p^2,(p+q)^2)p_\mu ~.
\label{corr}
\eq
Insertion of a complete set of hadronic states 
with $B$--meson quantum numbers between the 
currents in (\ref{corr}) entails relations between
the physical form factors $f^+$ and $f^+ +f^-$  
and the invariant amplitudes $F$ and $\widetilde{F}$, respectively. 
More definitely, for $\widetilde{F}$ one finds
\be
\widetilde{F}(p^2,(p+q)^2)=
\frac{m_B^2f_B (f^+(p^2)+f^-(p^2))}{m_b(m_B^2-(p+q)^2)}
+\int\limits_{s_0^h}^\infty ds \frac{\tilde{\rho}^h(p^2,s)}{s-(p+q)^2}~,
\label{hadr}
\ee
where the term proportional to $f^+ +f^-$ arises from the  
the contribution of the ground state $B$ meson, while the
integral over the  spectral density 
$\tilde{\rho}^h$ represents the contributions from  
excited resonances and continuum states above the threshold 
energy $\sqrt{s_0^h}$. 
In deriving this hadronic representation of $\widetilde{F}$
we have used the matrix element (\ref{def}) and
\be
\langle B \mid \bar{b}i\gamma_5d\mid 0\rangle=m_B^2f_B/m_b ~,
\label{decayconst}
\ee
$f_B$ being the $B$ meson decay constant.

In \cite{BKR,BBKR}, the invariant amplitude $F$ of the same
correlation function (\ref{corr}) is calculated
by expanding the $T$-product of 
the currents near the light-cone at $x^2 = 0$. 
The leading 
contribution to the operator product expansion (OPE)
is obtained by contracting 
the $b$-quark fields in (\ref{corr}) and inserting 
the free $b$-quark propagator
\bq
\langle 0|T\{b(x)\bar{b}(0)\}|0\rangle = i
\int \frac{d^4k}{(2\pi)^4}e^{-ikx}
\frac{\not\!k+m_b}{k^2-m_b^2}~.
\label{prop}
\eq
Substitution of (\ref{prop}) in (\ref{corr}) yields
\bq
F_\mu(p,q)=i\int \frac{
d^4x\,d^4k}{(2\pi )^4(m_b^2-k^2)}
e^{i(p-k)x}\Bigg( m_b
\langle \pi (q)|\bar{u}(x)\gamma_\mu \gamma_5 d(0)|0\rangle 
\nonumber
\\
+k^\nu \langle\pi(q) 
|\bar{u}(x)\gamma_\mu \gamma_\nu\gamma_5 d(0)|0\rangle\Bigg) ~.
\label{corr2}
\eq
This approximation is valid in the region of 
momenta $(p+q)^2 \ll m_b^2$ and 
\be
p^2 \leq m_b^2-2m_b\chi ~, 
\label{range}
\ee
$\chi$ being a $m_b$-independent scale
of order $\Lambda_{QCD}$. 
Since the pion is on-shell, 
$q^2= m_\pi^2$ vanishes in the chiral limit adopted throughout this 
calculation. The above restrictions ensure that the $b$ quark is sufficiently
off-shell, and that the resonances in the $\bar{u}b$ channel are sufficiently
far away.

The bilocal vacuum-to-pion matrix elements of light-quark fields 
encountered on the r.h.s. of (\ref{corr2}) are expanded around
$x^2 = 0$ leading to a series of contributions with increasing twist.
The coefficient functions of this expansion can be
parametrized by pion wave functions on
the light-cone \cite{BL,CZ,BF}. 
Including terms up to order $x^2$,
the light-cone expansion of the first matrix element 
in (\ref{corr2}) reads
\begin{eqnarray}
\langle\pi(q)|\bar{u}(x)\gamma_\mu\gamma_5d(0)|0\rangle =
-iq_\mu f_\pi\int\limits_0^1du\,e^{iuqx}
\left(\varphi_\pi (u)+x^2g_1(u)\right)
\nonumber \\
+
f_\pi\left( x_\mu -\frac{x^2q_\mu}{qx}\right)\int_0^1
du\,e^{iuqx}g_2(u) ~.
\label{phi}
\end{eqnarray}
Here, $\varphi_\pi$ is the leading twist~2 wave function, while $g_1$ 
and $g_2$ are twist~4 wave functions. Upon substitution of 
$\gamma_\mu \gamma_\nu = -i\sigma_{\mu\nu} + g_{\mu\nu} $, the
second term in (\ref{corr2}) is decomposed into 
the matrix elements 
\be
\langle\pi(q)\mid \bar{u}(x)i\gamma_5d(0)\mid 0\rangle=
f_\pi \mu_\pi \int_0^1du~e^{iuqx}\varphi_{p}(u)
\label{phip}
\ee
and 
\be
\langle\pi(q)\mid\bar{u}(x)\sigma_{\mu\nu}\gamma_5d(0)\mid 0\rangle=
i(q_\mu x_\nu -q_\nu x_\mu )\frac{f_\pi \mu_\pi}{6}
\int_0^1 du ~e^{iuqx}\varphi_{\sigma }(u) \, ,
\label{phisigma}
\ee
with $\mu_\pi=m_\pi^2/(m_u + m_d)$.
In leading order, these matrix elements involve  
the twist~3 wave functions $\varphi_p$ and $\varphi_\sigma$.
It is worth pointing out that the path-ordered gluon operator
\be
\Pi_G = P\exp \left\{ig_s \int\limits_0^1d\alpha~ 
        x_\mu A^\mu (\alpha x)\right \}
\label{factor}
\ee
ensuring gauge invariance of the above matrix elements 
is unity in the light-cone gauge, $x_\mu A^\mu=0$,
assumed here. Therefore, the factor $\Pi_G$ is not shown explicitly
in (\ref{phi})--(\ref{phisigma}).

Substitution of (\ref{phi})--(\ref{phisigma}) in
(\ref{corr2}), integration over $x$ and $k$,
and collection of  all terms 
proportional  to $p_\mu$ yield  
the following expression for  the 
invariant amplitude $\widetilde{F}$:
\bq
\widetilde{F}_{QCD}(p^2,(p+q)^2)=f_\pi \int\limits_0^1 \frac{du}
{m_b^2-(p+uq)^2}\left\{ 
\mu_\pi\varphi_{p}(u)
+\frac{\mu_\pi \varphi_{\sigma }(u) }{6u}
\right.
\nonumber
\\
\left. \times \Bigg( 1-
\frac{m_b^2-p^2}{m_b^2-(p+uq)^2}\Bigg)
+ \frac{2m_b g_2(u)}{m_b^2-(p+uq)^2} \right\} ~.
\label{Ftilde}
\eq
The index QCD has been added to distinguish the above  
representation of the invariant function $\widetilde{F}$ in terms of quark and
gluon degrees of freedom from the
hadronic representation given in (\ref{hadr}).
Note that the twist 2 and 4 wave functions $\varphi_\pi$ 
and $g_1$, respectively, do not contribute to $\widetilde{F}$.
This is 
obvious from the definition (\ref{phi}).
In general, the correlation function (\ref{corr}) also receives 
contributions from gluon emission by the $b$ quark.
This correction involves  quark-antiquark-gluon 
wave functions as described 
in \cite{BKR,BBKR}. However, 
direct calculation shows that up to the twist 3 and 4 the 
three-particle correction vanishes 
in the invariant amplitude $\widetilde{F}$.
Hence, to twist 4 accuracy,  the result for $\widetilde{F}$
turns out to be remarkably simple, at least when
compared with the corresponding expression for the invariant 
amplitude $F$ given in \cite{BKR,BBKR}.

The equality of the two representations  
(\ref{hadr}) and (\ref{Ftilde}) of $\widetilde{F}$
implies a sum rule for $f^+ + f^-$ which, however, is only useful
if one can remove the unknown contributions from the 
excited and continuum states. This is possible to a reasonable approximation
by making use of quark-hadron duality.
Following the standard procedure, the integral in (\ref{hadr})
over the hadronic spectral function above the ground state is replaced
by the corresponding integral over the imaginary part of 
$\widetilde{F}_{QCD}$. Formally, one can substitute 
\be
\tilde{\rho}^h(p^2,s)\Theta (s-s_0^h)=
\frac{1}{\pi}\mbox{Im} ~\widetilde{F}_{QCD}(p^2,s)\Theta (s-s_0^B)~,
\label{dual}
\ee
where $s_0^B$ is an effective threshold 
parameter separating the duality interval of the ground state from 
the one of the higher states. With this approximation, it is straightforward 
to subtract the contribution of the excited and continuum states 
from the basic equation 
given by (\ref{hadr}) and (\ref{Ftilde}).
After performing the obligatory Borel transformation in $(p+q)^2$, one finally arrives at the  
sum rule
\be
f_B(f^+(p^2)+f^-(p^2))=\frac{m_b}{\pi m^2_B}\int\limits_{m_b^2}^{s_0^B} 
\mbox{Im} \tilde{F}_{QCD}(p^2, s)
\exp\left(\frac{m^2_B-s}{M^2}\right)ds~,
\label{sr5}
\ee
$M^2$ being the Borel mass parameter.

The remaining task is then to derive $\mbox{Im} \widetilde{F}_{QCD}(p^2,s)$
from (\ref{Ftilde}). This is explained below.
Using $(p+uq)^2 =(1-u)p^2 +u(p+q)^2$ and changing   
variable from 
$u$ to $ s = (m_b^2-p^2)/u +p^2$  one can rewrite (\ref{Ftilde})
as follows:
\be
\widetilde{F}_{QCD}(p^2,(p+q)^2) = 
\sum_{i=1,2}\int^\infty_{m_b^2} ds 
\frac{\rho_i(p^2,s)}{(s-(p+q)^2)^i} ~,
\label{dispers}
\ee
where 
\be
\rho_1(p^2,s)=\frac{f_\pi \mu_\pi}{s-p^2}\left( 
\varphi_{p}(u)
+\frac{\varphi_{\sigma }(u) }{6u}
\right)~,
\label{rhoone}
\ee
\be
\rho_2(p^2,s)=
\frac{f_\pi}{m_b^2-p^2}\left(-\frac{\mu_\pi\varphi_\sigma(u) }{6u}
(m_b^2-p^2) + 2m_b g_2(u) \right) ~.
\label{rhorho}
\ee
The term $i=1$ in (\ref{dispers}) already
has the form of a dispersion integral in the variable $(p+q)^2$. 
In order to achieve this also for the term
$i=2$ one has to perform a partial integration yielding in total:
\bq
\widetilde{F}_{QCD}(p^2,(p+q)^2)& = & 
\int^\infty_{m_b^2}  
\frac{ds}{s-(p+q)^2}~\left(\rho_1(p^2,s) + \frac{d\rho_2(p^2,s)}{ds}\right)
\nonumber
\\
&&- \int^\infty_{m_b^2}ds \frac{d}{ds}\left( \frac{\rho_2(p^2,s)}
{s-(p+q)^2}\right)~.
\label{disp1}
\eq 
Since the wave functions $\varphi_\sigma$ and $g_2$ 
vanish at $u=0$ and $u=1$,
that is $s=\infty$ and $s=m^2_b$, respectively,
as can be seen from the explicit expressions given in the subsequent
section,
the second integral in (\ref{disp1}) is zero.
Hence, the imaginary part of $\widetilde{F}_{QCD}$
can be directly read off from the integrand of the first integral:
\be
\frac1{\pi}\mbox{Im}\widetilde{F}_{QCD}(p^2,s) = 
\rho_1(p^2,s) + \frac{d\rho_2(p^2,s)}{ds}        ~.
\label{Im}
\ee
Substitution of (\ref{Im}) in  (\ref{sr5}) yields
\bq
f_B(f^+(p^2)+f^-(p^2))=
\frac{m_b}{m^2_B}\int\limits_{m_b^2}^{s_0^B} 
\left(\rho_1(p^2,s) + \frac{d\rho_2(p^2,s)}{ds}\right)
\exp\left(\frac{m^2_B-s}{M^2}\right)ds
\nonumber
\\
=\frac{m_b}{m^2_B}\left\{ \int\limits_{m_b^2}^{s_0^B}
\left(\rho_1(p^2,s) + \frac{\rho_2(p^2,s)}{M^2}\right)
\exp\left(\frac{m^2_B-s}{M^2}\right)ds+
\rho_2(p^2,s_0^B)\exp\left(\frac{m^2_B-s_0^B}{M^2}\right)\right\}.
\label{dispers2}
\eq
In previous applications of QCD light-cone sum rules 
with the exception of
the recent calculation of the $B\rightarrow \rho$ form factors in \cite{BB},
surface terms similar to the last
term on the r.h.s. of (\ref{dispers2}) have been neglected.
They originate from higher twist contributions to 
Im$\widetilde{F}_{QCD}$ 
and play a minor role numerically.
Nevertheless, in order to subtract the contributions from
excited and continuum states in the   
duality approximation consistently we take these terms 
into account in the present calculation. 
 
The final sum rule for $f^+ + f^-$ follows from (\ref{rhoone}), (\ref{rhorho})
and (\ref{dispers2}) after returning to the variable $u$:
$$
f_B(f^+(p^2) + f^-(p^2))=\frac{f_\pi \mu_\pi m_b}{m_B^2 }\exp\left(
\frac{m_B^2}{M^2}\right)\Bigg\{
\int\limits^{1}_{\Delta} 
\frac{du}{u}
\exp \left( - \frac{m_b^2-p^2(1-u)}{u M^2}\right)
$$
$$
\times\Bigg( 
\varphi_{p}(u) 
+ \frac{\varphi_\sigma(u)}{6u} 
\left(1- 
\frac{m_b^2-p^2}{uM^2}\right) 
+ \frac{2m_bg_2(u)}{\mu_\pi uM^2}\Bigg)
$$
\be
+\exp\left(- \frac{s_0^B}{M^2}\right)
\left(- \frac{\varphi_\sigma(\Delta)}{6\Delta} 
+ \frac{2m_bg_2(\Delta)}{\mu_\pi(m_b^2-p^2)}\right)\Bigg\}
\label{fplusminus}
\ee
with $\Delta = (m_b^2-p^2)/(s_0^B-p^2) $.
For comparison and later use, we also quote the analogous sum rule for $f^+$
obtained in \cite{BKR,BBKR}:
$$
f_B f^+( p^2)=\frac{f_{\pi} m_b^2}{2 m_B^2}
\exp \left( \frac{m_B^2}{M^2}\right)
\Bigg\{
\int\limits_\Delta^1\frac{du}{u} 
\exp\left(-\frac{m_b^2-p^2(1-u)}{uM^2} \right)
$$
$$
\times\Bigg( \varphi_\pi(u) 
+ \frac{\mu_\pi}{m_b}
\left[u\varphi_{p}(u) + \frac{
\varphi_{\sigma }(u)}{3}\left(1 + \frac{m_b^2+p^2}{2uM^2}\right) \right]
- \frac{4m_b^2 g_1(u)}{ u^2M^4 }
$$
\be
+\frac{2}{uM^2} \int^u_0 g_2(v)dv
\left(1+ \frac{m_b^2+p^2}{uM^2} \right ) \Bigg)+ t^+(s_0^B,p^2,M^2)
+f^+_G(p^2,M^2) \Bigg\}~.
\label{fplus}
\ee
Here, we have added the surface term
$t^+$ which was neglected previously, and denoted
the contribution from the quark-antiquark-gluon
wave functions of twist~3 and 4 by
$f^+_G$. 
The explicit expressions for $t^+$ and $f^+_G$ can be found in the Appendix. 

Since very recently, the perturbative $O(\alpha_s)$ correction 
to the leading twist 2 piece of the light-cone sum rule 
(\ref{fplus}) for $f^+$ 
is also known \cite{KRWY,Bagan}. However, the corresponding QCD corrections
to the twist 3 term in (\ref{fplus}) as well as in the sum rule
(\ref{fplusminus}) for $f^+ + f^-$ still remain to be calculated.
Hence, for consistency, we will not include the $O(\alpha_s)$ effects 
in $f^+$ ¤in the present analysis.

\section{Numerical results}

For the numerical analysis of the new sum rule 
(\ref{fplusminus}) we use the same input as in the evaluation 
of the sum rule (\ref{fplus}) in \cite{BKR,BBKR}. From experiment we take
$f_\pi = 132$ MeV and $m_B=5.279$ GeV, whereas the parameters 
$m_b=4.7 \pm 0.1 $ GeV, $s_0^B= 35 \mp 2$ GeV$^2$, and $f_B=140 \mp 30 $ 
MeV are extracted from the QCD sum rule for the correlator of two
$\bar{b}\gamma_5 u$ currents. For consistency, the $O(\alpha_s)$ 
correction is not included in the latter two-point sum rule.
This is reflected by the low value of $f_B$.
Because of cancellations of QCD corrections in the ratios of 
(\ref{fplusminus}) and $f_B$, respectively, (\ref{fplus}) and $f_B$,
the remaining corrections to the form factors themselves may in fact
be small. This is precisely what happens in the case of 
$f^+$ as  has been shown in \cite{KRWY,Bagan}.  
 
Furthermore, the explicit expressions for the pion
wave functions up to twist 4 are collected in \cite{BBKR}. 
Those entering the new sum rule (\ref{fplusminus}) 
are given below for completeness:
\begin{eqnarray}
\varphi_\pi(u,\mu) &=& 6 u\bar{u}\Big[1+a_2(\mu)
\frac{3}2[5(u-\bar{u})^2-1]+
 a_4(\mu)\frac{15}8[21(u-\bar{u})^4-14(u-\bar{u})^2+1]\Big]\,,
\label{twist2wf}
\\
\varphi_p(u,\mu) &=& 1+B_2(\mu)\frac12\left[3(u-\bar{u})^2-1\right]
+B_4(\mu)\frac18\left[35(u-\bar{u})^4
-30(u-\bar{u})^2+3\right]\,,
\label{bbb}
\\
\varphi_\sigma (u,\mu) &=& 6u\bar{u}\Big[ 
1+C_2(\mu)\frac{3}2\left(5(u-\bar{u})^2-1\right)
+C_4(\mu)\frac{15}8(21(u-\bar{u})^4
\nonumber
\\
&&{}-14(u-\bar{u})^2+1)\Big]\,,
\label{tw3}
\\
g_1(u,\mu) &=& \frac{5}2\delta^2(\mu)\bar{u}^2u^2+\frac{1}{2}\varepsilon(\mu)
\delta^2(\mu)[\bar{u}u(2+13\bar{u}u)+10u^3\ln u(2-3u+\frac65u^2)
\nonumber
\\
&&{}+10\bar{u}^3\ln \bar{u}(2-3\bar{u}+\frac65\bar{u}^2)]\,,
\label{tw4}
\\
g_2(u,\mu) &=& \frac{10}3\delta^2(\mu)\bar{u}u(u-\bar{u})\,,
\label{tw4a}
\end{eqnarray}
with $\bar{u}=1-u$ and $\mu$ being the renormalization scale.
For a detailed discussion of the wave functions we refer the reader
to the original literature \cite{BL,CZ,BF}.
Recent reviews and references can be found in \cite{review,braunreview}.  
Here, the specification of the various coefficients together with a few 
comments may suffice.
The terms proportional to the coefficients $a_i$, $B_i$ and $C_i$ 
represent  scale-dependent nonasymptotic corrections. 
They vanish logarithmically 
as $\mu \rightarrow \infty$.

In leading--order, the value of the scale $\mu$ is ambiguous.   
As a reasonable choice we take
$\mu_b =\sqrt{m_B^2-m_b^2}=2.4~\rm{GeV}$.
With this choice, 
the estimate in \cite{BBKR} gives 
$a_2(\mu_b)=0.35$, $a_4(\mu_b)= 0.18$, $B_2(\mu_b)=0.29$, $B_4(\mu_b)=0.58$, 
$C_2(\mu_b)=0.059$, $C_4(\mu_b)=0.034$,
$\delta^2(\mu_b)=0.17$ GeV$^2$, and $\varepsilon(\mu_b)=0.36$. 
Furthermore, the PCAC relation 
$f_\pi^2 \mu_\pi(\mu) = -2\langle \bar{q}q\rangle(\mu)$
and $\langle \bar{q}q \rangle(\mu_b)= 
(-260 \pm 10~\mbox{MeV})^3$ can be used to fix the final parameter 
appearing in (\ref{fplusminus}) and (\ref{fplus}), namely
$\mu_\pi = \mu_\pi(\mu_b) = 2.0 \pm 0.25$ GeV. 
 
In the case of the sum rule (\ref{fplus}) for $f^+$ 
the acceptable range of values of the 
Borel parameter $M^2$ was found to be $8 < M^2 <12 $ GeV$^2$ \cite{BBKR}.
In the sum rule (\ref{fplusminus}) for $f^+ +f^-$ we take the same
interval after having checked that in this range of $M^2$ 
the twist 4 contribution does not exceed 10\%, 
and that the excited and continuum states 
(\ref{dual}) do not contribute more than 30\%. These are
the usual conditions posed in  sum rule applications. 
 
Having specified the necessary numerical input, we are now ready to present
quantitative results. In Fig. 1, the sum rule (\ref{fplusminus})
is plotted as a 
function of Borel parameter $M^2$. One can see that the variation
is very moderate in the range $8 < M^2 <12 $ GeV$^2$, at least for 
$p^2 \leq 15$ GeV$^2$. This also applies to the sum rule (\ref{fplus}).
However,
at $p^2 > 17$ GeV$^2$ the dependence on $M^2$ becomes strong,
indicating 
that one is getting too close to the physical states in this channel.
Fig. 2 shows the
momentum dependence of the form factors $f^+(p^2)+f^-(p^2)$,
$f^+(p^2)$, and 
$f^0(p^2)$ for the central value $M^2= 10$ GeV$^2$. 
Here, the scalar form factor $f^0$ is calculated from the other two
form factors using the relation (\ref{f0}). In particular,
at zero momentum transfer we predict
\bq
f^+(0) = f^0(0) =0.30,
\label{fpl}
\\
f^+(0) + f^-(0)= 0.06.
\label{fplmin}
\eq

One of the greatest virtues of the sum rule approach is the possibility
to estimate the theoretical uncertainties in the predictions, at least
in principle. In practice, this task is not straightforward and, hence,
the estimates should be considered with caution.
The uncertainties in (\ref{fplus}) and 
(\ref{fplusminus}) induced by the input parameters $M^2$, $m_b$, $s_0^B$,
$f_B$, and $\mu_{\pi}$ can be investigated 
by varying these parameters simultaneously 
in the sum rules for the form factors and for $f_B$. For instance,
if the Borel mass is allowed to vary within the interval quoted above, 
$f^0$ is found to deviate by $\pm 3\%$ ($\pm 5\%$) 
at small (large) $p^2$ from the value obtained with the nominal
choice $M^2= 10$ GeV$^2$. The corresponding uncertainty in $f^+ + f^-$
can be inferred from Fig. 1. Furthermore, 
if $m_b$ and $s_0^B$ are varied in a correlated way 
within the ranges given at the beginning of this section
such that one achieves  
maximum stability of the sum rule for $f_B$, $f^0$ 
changes by about $\pm 2\%$ relative to
the value obtained with the nominal choice of parameters. 
This uncertainty is $p^2$-independent.     

Another source of uncertainty is the precise shape of the
pion wave functions.
Keeping all other parameters fixed, we have studied 
the sensitivity of our results to the 
nonasymptotic terms in the twist 2 and 3 wave functions 
given in (\ref{twist2wf})--(\ref{tw3}).
In Fig. 3 we compare the prediction on $f^0$ with the coefficients
$a_i$, $B_i$, and $C_i$ $(i=2,4)$ as given earlier in this section 
to the result obtained by putting them to zero.
The shifts are momentum dependent reaching 
-10 \% at $p^2 = 0$ and +5 \% at 15 GeV$^2$. The real uncertainty 
is certainly less than that.

In addition, there is some uncertainty due to the 
truncation of the light-cone expansion or, in other words, due to
the neglect of terms with twist larger than 4. An upper limit 
on this uncertainty should be set by the 
size of the twist 4 contribution. The influence of the latter on 
the scalar form factor $f^0$
is displayed in Fig. 4. The twist 4 terms decrease $f^0$ by    
2 \% at small $p^2$ and  
by 5 \% at large $p^2$.

Finally, the present lack of knowledge of the perturbative QCD 
corrections to the twist 3 contributions in the sum rules
(\ref{fplus}) and (\ref{fplusminus}) gives rise to uncertainties
which will eventually be removed in near future. The scalar form factor
$f^0$ is concerned in particular at large momentum transfer where twist 3
should dominate as expected from the relation (\ref{f0}). Conversely,
at small $p^2$ $f^0$ coincides with the form factor $f^+$ which receives
the leading contribution from the twist 2 wave function $\varphi_{\pi}$. 
The $O(\alpha_s)$  corrections to this piece are known. They change
the lowest order estimate (\ref{fpl})
by only 10 \% \cite{KRWY}:
\bq
f^+(0) = f^0(0)= 0.27 ~.
\label{fplNLO}
\eq
As already pointed out, this is due 
to a remarkable cancellation of the corrections 
to the sum rule (\ref{fplus}) for $f^+f_B$ and the  
sum rule for $f_B$ in their ratio. 
Whether or not a similar cancellation takes place
at the twist 3 level ist questionable and can only be decided by 
direct calculation. For the time being the total uncertainty in $f^0$
is estimated to be about 20 \% at small $p^2$ and 30 \% at large $p^2$.

The sum rules (\ref{fplusminus}) and 
(\ref{fplus}) for the $ B\rightarrow \pi$ form factors 
are easily converted into sum rules for 
the corresponding $D\rightarrow \pi$  form factors. Formally, one
only has to replace the flavour indices 
$b$ by $c$, and $B$ by $D$. 
Because of the relatively light charm mass,
the region of validity of the sum rules covers only the low
momentum region $0\leq p^2\leq 1.0$ GeV$^2$
of the kinematically allowed range of $p^2$. 
The values of the input parameters are
$m_c=1.3 \pm 0.1$ GeV, $s_0^D=6 \mp 1$ GeV$^2$, and
$f_D=170 \pm 20$ MeV. The scale $\mu$ is taken to be
$\mu_c = \sqrt{m_D^2-m_c^2} = 1.3$ GeV, while the fiducial range of the 
Borel mass is $3 ~\mbox{GeV}^2 < M^2 < 5 ~\mbox{GeV}^2$.
Correspondingly, the value of $\mu_{\pi}$ is lowered to
$\mu_\pi(\mu_c) = 1.8 \pm 0.5$ GeV. The numerical values of the 
nonasymptotic coefficients in the pion wave functions at the scale $\mu_c$ 
are given in \cite{BBKR,review}. There, one can also find further
remarks on the above choice of parameters. 
Our numerical predictions for the $D\rightarrow \pi$
form factors are illustrated in Fig. 5 using the central values
for the various input parameters. 
At zero momentum transfer, we estimate 
\bq
f^{+}(0) = f^0(0) = 0.68 ~,
\label{Dfpl}
\\
f^{+}(0) + f^{-}(0) = 0.52~.
\label{Dfplmin}
\eq

\section{Predictions for $B \to \pi \bar{\tau} \nu_\tau$}

The $B \to \pi$ form factors can be measured most directly
in the weak semileptonic decays 
$B \to \pi \bar{l} \nu_l$ where $ l = e,\mu$ or $\tau$.
The distribution of the momentum transfer squared 
in these decays is given by  
$$
\frac{d\Gamma}{dp^2} = 
\frac{G_F^2|V_{ub}|^2}{24\pi^3}
\frac{(p^2-m_l^2)^2 \sqrt{E_\pi^2-m_\pi^2}}{p^4m_B^2}
\Bigg\{ \left(1+\frac{m_l^2}{2p^2}\right)
m_B^2(E_\pi^2-m_\pi^2)\left[f^+(p^2)\right]^2 
$$
\be
+ \frac{3m_l^2}{8p^2}(m_B^2-m_\pi^2)^2
\left[f^0(p^2)\right]^2\Bigg\}
\label{dG}
\ee
with $E_\pi= (m_B^2+m_\pi^2 -p^2)/2m_B$ being the pion energy in the 
$B$ rest frame. Another interesting observable is the 
distribution of the charged lepton energy $E_l$ in the $B$ rest frame:
\bq
\frac{d\Gamma}{dE_l}&=&
\frac{G_F^2|V_{ub}|^2}{64\pi^3}\int^{p_{max}^2}_{p_{min}^2} dp^2
\Bigg\{ \Bigg[ 8\frac{E_l}{m_B}(m_B^2-m_\pi^2+p^2)
\nonumber
\\
&&-4(p^2+4E_l^2) +\frac{m_l^2}{m_B^2}\left(8m_BE_l-3p^2+4m_\pi^2
\right ) -\frac{m_l^4}{m_B^2}\Bigg] \left[f^+(p^2) \right]^2
\nonumber
\\
&&+\frac{2m_l^2}{m_B^2}\Bigg[2m_B^2+p^2-2m_\pi^2-4m_BE_l+m_l^2 \Bigg]
f^+(p^2)f^-(p^2) 
\nonumber
\\
&&+\frac{m_l^2}{m_B^2}(p^2-m_l^2)\left[f^-(p^2) \right]^2 \Bigg\}
\label{spectrum}
\eq
with $p^2_{{max}\atop{min}}= m_B(E_l\pm \sqrt{E_l^2-m_l^2})+ O(m_\pi^2)$.
Although the form factors are calculated in the chiral limit, 
otherwise the finite pion mass is taken into account.
In the case of light leptons $l=e,\mu$ 
the form factor $f^0$ or, equivalently, $f^-$ plays a negligible
role because of the smallness of the electron and muon masses.
Hence, these decay modes can provide information only on the form factor
$f^+$. In contrast, the decay  $B \to \pi \bar{\tau} \nu_\tau$ 
is also sensitive to the form factor $f^0$. We therefore concentrate
here on the latter case. 

The sum rule results described in the preceding two sections allow to predict
the decay spectra in the momentum region $0 \leq p^2 \leq 17$ GeV$^2$.
In order to include higher momentum transfers and to predict integrated
widths one has to find another way to calculate the form factors 
up to the kinematical endpoint $p^2=(m_B-m_\pi)^2 = 26.4$ GeV$^2$.
In \cite{BBKR}, the single-pole approximation
\be
f^+(p^2)= \frac{f_{B^*}g_{B^*B\pi}}{2m_{B^*}(1-p^2/m_{B^*}^2)}
\label{onepole}
\ee
was used. 
Since the vector $B$ ground state is only about 50 MeV heavier than the
pseudoscalar $B$ , the
$B^*$ pole is very near to the endpoint region. Consequently,
at maximum $p^2$ the single-pole approximation can be
expected to be very good. Moreover, the
strong $B^*B\pi$ coupling which determines the normalization of the
form factor at large $p^2$ can be calculated from the same 
correlation function
(\ref{corr}) from which the sum rule (\ref{fplus}) for $f^+$
at low to intermediate values of $p^2$ is derived. To this end 
one employs a double dispersion relation. The method and
results are described in \cite{BBKR} and reviewed in \cite{review}.
Extrapolation of the single-pole model to smaller $p^2$ matches
quite well with the direct estimate from
the light-cone sum rule (\ref{fplus})
at intermediate momentum transfer
$p^2=15$ to $20$ GeV$^2$. This provides us with a consistent and 
complete theoretical prediction of $f^+$. 

Unfortunately, it is doubtful that a similar procedure 
can be applied to the scalar form factor $f^0$, 
because the scalar $B$ ground state is 
expected to be about 500 MeV heavier than the pseudoscalar $B$.
Thus, the scalar $B$ pole may be too distant 
from the kinematical endpoint of the $B \to \pi$ transition
for the single-pole approximation to hold.
Nearby nonresonant $B\pi$ states and excited scalar 
resonances may give comparable contributions.

Interestingly, there exists 
a model-independent 
constraint on the behaviour of the form factor $f^0$ at large 
$p^2 \simeq m_B^2$ , i.e., near the kinematical endpoint.
The constraint \cite{Vol,DKS} is derived from a Callan-Treiman type relation 
obtained by combining current algebra and PCAC: 
\be
\mbox{lim}_{~p^2\rightarrow m_B^2}~f^0(p^2) = f_B/f_\pi~.
\label{CT}
\ee
In the following we make use of this bound 
in order to illustrate the sensitivity of the decay spectra in
$B \to \pi \bar{\tau} \nu_\tau$ to the scalar form factor.

The form factor $f^0$ is extrapolated linearly from 
the value at $p^2=15$ GeV$^2$ where the sum rules (\ref{fplusminus})
and (\ref{fplus}) still hold to the value at $p^2 \simeq m_B^2$ 
dictated by (\ref{CT}). To be conservative we take 
$f_B= 150$ to 210 MeV in accordance with 
recent lattice data \cite{Flynn} and with QCD sum rule estimates
(the latter including the perturbative correction, see e.g. \cite{KRWY}).
This is shown in Fig. 6 together with lattice estimates of $f^0$.
Obviously, the lattice data favour the lower extrapolation. 
The distributions of $p^2$ and $E_\tau$ in $B \to \pi \bar{\tau} \nu_\tau$ 
resulting from the upper and lower bounds on $f^0$
are plotted in Fig. 7 and Fig. 8, respectively.
This study demonstrates that  
measurements of these decay spectra at $B$ factories should provide 
interesting information on the elusive form factor $f^0$.

For the integrated partial width we predict 
\be
\Gamma(B^0\rightarrow\pi^- \tau^+ \nu_\tau) = 
5.7 ~\mbox{to} ~6.5 ~|V_{ub}|^2 ~\mbox{ps}^{-1} ~,
\label{tau}
\ee
where the range corresponds to
the two extrapolations considered in Figs. 7 and 8.
The theoretical uncertainties in the sum rule calculations discussed 
in sect. 3 are not included.
The latter drop out to a large extent in the ratio
\be
\frac{\Gamma(B^0\rightarrow\pi^- \tau^+ \nu_\tau)}
{\Gamma(B^0\rightarrow\pi^- e^+ \nu_e)}= 0.75 ~\mbox{to} ~0.85 ~.
\label{r}
\ee
It should be noted that 
only the numerator is influenced by the scalar form factor.

The form factor $f^0$ also plays an important role 
in nonleptonic $B$ decays where it enters the factorized two-body
amplitudes for $B \rightarrow \pi h$. Depending on the mass of the meson h
these decays probe $f^0$ in the range $m_{\pi}^2 < p^2 < m_{\psi}^2$.  
This is similar in $D$ decays. However, there 
the form factor $f^0$ cannot be measured independently in semileptonic decays
because only the electron and muon modes are kinematically accessible.

\section{Dependence on the heavy quark mass}

The light-cone sum rules (\ref{fplusminus}) and (\ref{fplus})
offer the possibility to systematically
investigate the dependence of heavy-to-light form factors 
on the heavy quark mass. Using the familiar scaling
relations for mass parameters and decay constants, to wit
\be
m_B = m_b+\bar{\Lambda}~,~~~ s_0^B = m_b^2 + 2m_b\omega_0 ~,
~~~M^2= 2m_b\tau~,
\label{hqet}
\ee
\be
f_B = \hat{f}_B/\sqrt{m_b}~, 
\label{fBhat}
\ee
where in the heavy quark limit 
$\bar{\Lambda}$, $\omega_0$, $\tau$, $\hat{f}_B$ are
$m_b$-independent quantities, it is rather straightforward to
expand the sum rules in inverse powers of $m_b$. 
At $p^2 = 0$, the leading terms are given by 
$$
f^+(0) = m_b^{-3/2}\frac{f_\pi}{2\hat{f}_B}
\exp\left( \frac{\bar{\Lambda }}{\tau }\right)
\Bigg\{\int^{2\omega _0}_0 d \rho ~
\exp\left[-\frac{\rho }{2\tau }\right]
$$
$$
\times\left( -\rho\varphi_\pi'(1)
+\mu_\pi\left[\varphi_p(1)-
\frac{\rho}{12\tau}\varphi_\sigma'(1) 
\right]\right) 
$$
\be
-\frac{\mu_\pi \omega_0}6 \exp \left( -\frac{\omega_0}{\tau} \right)
\varphi_\sigma'(1)\Bigg\}
+O(m_b^{-5/2}) ~,
\label{zerolimit}
\ee
and 
$$
f^+(0) +f^-(0) = m_b^{-3/2}
\frac{f_\pi\mu_\pi}{\hat{f}_B}
~\exp\left( \frac{\bar{\Lambda }}{\tau }\right)
\Bigg\{\int^{2\omega _0}_0 d \rho~
\exp\left[-\frac{\rho }{2\tau }\right]
$$
\be
\times \left[\varphi_p(1) +\frac{\rho}{12\tau}\varphi_\sigma'(1) \right] 
+\frac{\omega_0}3 \exp\left(-\frac{\omega_0}{\tau}\right)
\varphi_\sigma'(1)\Bigg\}
+O(m_b^{-5/2}) ~,
\label{zerolimit1}
\ee
where $ \varphi'_a(1)$ stands for the derivative 
$ d\varphi_a/du$ at the endpoint $u=1$.
It is important to note that whereas the twist 2 and 3 
two-particle wave functions 
survive in the asymptotic limit  
(\ref{zerolimit}) and (\ref{zerolimit1}), the higher-twist and
three-particle components are suppressed by one additional power
of $m_b$, and die out.

The asymptotic scaling laws derived above can be understood as follows. 
At $p^2 = 0$, the integration region $\Delta = m_b^2/s_0^B \leq u \leq 1$
in (\ref{fplusminus}) and (\ref{fplus}) is rather narrow. 
In fact, it vanishes in the infinite
mass limit as $1 - m_b^2/s_0^B \sim 2\omega_0 /m_b$.
In this limit, the asymptotic twist 2 and twist 3 wave functions
behave like 
\be
\varphi_\pi \sim \varphi_\sigma \sim (1-u) \sim \omega_0 /m_b ~\mbox{and}
~\varphi_p \sim 1~. 
\label{scaling}
\ee
Taking into account the extra factor $1/m_b$ multiplying $\varphi_p$,
and noticing that the factors $1/m_b$ times bracket multiplying 
$\varphi_\sigma$ in the sum rules approach unity at $m_b \rightarrow \infty$,
one sees that the twist 2 and 3 terms lead to
the same asymptotic scaling behaviour. The latter is determined   
by a factor $m_b^{-1}$ from the integrand, a factor $m_b^{-1}$ from the
integration region, and a factor $m_b^{1/2}$ from $1/f_B$.
The fact that the light-cone sum rule
predicts $f^+(0) \sim m_b^{-3/2}$  
was first noticed in \cite{CZ1}. 

The situation changes drastically when the momentum transfer becomes large
of order $m_b^2$.   
To be definite, at the boundary $p^2 = m_b^2-2m_b\chi$ considered in 
(\ref{range}), the integration region 
in (\ref{fplusminus}) and (\ref{fplus}) is finite and independent
of $m_b$. Therefore, the asymptotic scaling laws are simply determined
by the factors in front of the duality integrals in the sum rules, that
is $1/f_B \sim m_b^{1/2}$ in the case of $f^+$ 
and $1/m_bf_B \sim m_b^{-1/2}$ for $f^+ + f^-$.
Explicitly, one obtains
$$
f^+(p^2 = m_b^2-2m_b\chi)
\sim m_b^{1/2}\frac{f_\pi}{2\hat{f}_B}
\exp\left(\frac{\bar{\Lambda}}{\tau}\right)
\Bigg\{\int^1_{\Delta}\frac{du}{u}
~\exp\left[-\frac{\chi(1-u)}{\tau u}\right]
$$
$$
\times \left[ \varphi_\pi(u) 
+ \frac{\mu_\pi}{6u\tau}\varphi_{\sigma}(u)
- \frac1{u^2\tau^2}\left( g_1(u) - \int^u_0 g_2(v) dv\right ) \right]
$$ 
$$
+\frac1{\chi}\exp\left( -\frac{\omega_0}{\tau}
\right)
\Bigg[\frac{\mu_\pi}6 \varphi_{\sigma}(\Delta)-
\frac1{\chi}\left( 1+\frac{\chi+\omega_0}{\tau}
\right)\left( g_1(\Delta) -\int^\Delta_0 g_2(v)dv\right)
$$
\be
+ \frac1{\chi+\omega_0} \left( \frac{dg_1(\Delta)}{du}
- g_2(\Delta) \right)\Bigg] \Bigg\}
+O(m_b^{-1/2})~,
\label{mblimit}
\ee
and 
$$
[f^+ + f^-](p^2\sim m_b^2-2m_b\chi)
\sim m_b^{-1/2}\frac{f_\pi\mu_\pi}{\hat{f}_B}
~\exp\left(\frac{\bar{\Lambda}}{\tau}\right)\Bigg\{
\int^1_{\Delta}\frac{du}{u}
\exp\left[-\frac{\chi(1-u)}{\tau u}\right]
$$
$$
\times
\left[ \varphi_p(u) + \frac{\varphi_{\sigma}(u)}{6u}
\left(1-\frac{\chi}{u\tau}\right) 
+ \frac{g_2(u)}{\mu_\pi u\tau}\right]
$$
\be
+ \frac1{\chi}\exp\left(-\frac{\omega_0}{\tau}\right)
\left( -\frac16(\chi+\omega_0)\varphi_{\sigma}(\Delta) +
\frac{g_2(\Delta)}{\mu_\pi}\right) \Bigg\}
+O(m_b^{-3/2})~,
\label{mblimit1}
\ee
where $\Delta = \chi/ (\chi +\omega_0)$. In contrast to the
heavy quark limit at $p^2 = 0$, here also twist 4 contributes 
asymptotically. However, the contributions from three-particle
wave functions are still suppressed by an extra power of 
$1/m_b$.

Finally, using the relation (\ref{f0}) 
it is easy to check that the asymptotic scaling law of the 
form factor $f^0$ coincides with the expectation (\ref{zerolimit})
for $f^+$ at small $p^2$, and with (\ref{mblimit1}) for
$f^+ +f^-$ at large $p^2$. 

The above analysis shows   
that the light-cone expansion
in terms of wave functions with increasing twist is 
consistent with the heavy mass expansion.
The higher-twist contributions either 
scale with the same power of $m_b$ as the leading-twist term,
or they are suppressed by extra powers of $m_b$.
The sum rules nicely reproduce
the asymptotic dependence of the form factors $f^{\pm}$
on the heavy quark mass as derived in \cite{Vol,IW}
for small pion momentum in the rest frame of the $B$ meson.
In addition, they also allow to investigate the case
of large pion momentum where neither HQET nor the single-pole 
model can be trusted. 
Cleary, as  $p^2 \rightarrow 0$
excited and continuum states are expected to become
more and more important thus leading to a break-down of the single-pole
approximation. The change in the asymptotic mass dependence
of the light-cone sum rules when going from large to small 
momentum transfers can be considered as a signal of this break-down.
Claims in the literature which
differ from (\ref{zerolimit}) and (\ref{zerolimit1}) are 
often based on the pole model and therefore 
incorrect in our opinion. The above conclusions corroborate
similar analyses carried out for the 
$ B \to K^*$ \cite {ABS} and $B\to \rho$ \cite{BB} transition form factors.

In order to clarify the relevance of the asymptotic scaling behaviour 
in the mass range between $m_c$ and $m_b$, 
we have   studied  the functional dependence of the sum rules 
(\ref{fplusminus}) and (\ref{fplus}) on the heavy quark mass $m_Q$ 
numerically.  
The dependence of the parameters $m_B$, $s_0^B$ and $M^2$ on 
$m_Q$ is described approximately by the relations (\ref{hqet})
using $\bar{\Lambda}= 0.6$ GeV,
$\omega_0=1.4$ GeV, $\tau = 1.1$ GeV. Together with $m_Q=4.7$ GeV
this choice reproduces
the central values of $m_B$ etc. given at the beginning of sect. 3.
In order to consistently include the deviations of the decay constant
$f_B$ from the asymptotic scaling law (\ref{fBhat})
we have substituted $f_B$ by the corresponding two-point sum rule 
using again the relations (\ref{hqet}) analogously to the procedure
described above. The logarithmic mass dependence 
of the wave functions and vacuum condensates through the scale 
$\mu_Q = \sqrt{2m_Q\bar{\Lambda}}$  is taken into account. 
Fig. 9 shows the form factor $f^+(0)$
multiplied by the leading power $m_Q^{3/2}$ as a function of $1/m_Q$. 
Even at $m_Q>m_b$ there is 
still no sign that one is approaching the asymptotic limit.
On the contrary, the mass dependence of the nonasymptotic terms in the pion 
wave function becomes more and more important, at least for an 
intermediate mass range. At $m_Q <m_b$, the integration region is still
big enough to wash out these effects since the wave functions are 
normalized. In the region between $m_c$ and $m_b$, the mass-dependence
can be fitted to the following quadratic polynomial in $1/m_Q$:
\be
f^+(0)m_Q^{3/2}=3.3\mbox{GeV}^{3/2}\left( 1-\frac{1.5\mbox{GeV}}{m_Q}+
\frac{0.75\mbox{GeV}^2}{m_Q^2} \right),  
\label{fitt}
\ee 
indicating the existence of large $1/m_Q$ corrections in the physical
mass range. Similar results have been obtained in the case of $B\to K^*$ 
\cite{ABS} and $B\to \rho$ \cite{BB} form factors .

\section{Conclusion}

In this paper, we have derived a new QCD sum rule
for the combination $f^+(p^2) + f^-(p^2)$ of $B \rightarrow \pi$ 
(and $D\rightarrow\pi$) form factors. The light-cone approach 
used is designed for heavy-to-light transitions and incorporates
the nonperturbative dynamics in terms of light-cone wave functions
of the pion. The sum rule is essentially determined by the twist 3 
$q$-$\bar{q}$ wave functions. Terms involving the leading 
twist 2 wave function and the $q$-$\bar{q}$-$g$ wave function
of twist 3 and 4 are absent, while the twist 4 $q$-$\bar{q}$ wave functions 
give only small contributions. Higher-twist components are neglected. 
The sum rule is valid in the range of momentum transfers 
$0 \leq p^2 \leq 17$ GeV$^2$.

Combining the new result on $f^+ + f^-$ with the corresponding
calculation of $f^+$ in \cite{BBKR} we have been able to predict the
scalar form factor $f^0$.
This prediction is compared with recent lattice results \cite{Flynn}
in Fig. 6 and with quark model \cite{BSW} and different sum rule \cite{ColSan}
estimates in Fig. 10.
Within the inherent uncertainties of both approaches there is agreement
with the lattice results. However, the latter tend to be 
systematically lower than our prediction. This could be an indication
for the presence of perturbative QCD corrections which still need to be
calculated. Our result on $f^0$ also agrees with the 
lattice-constrained parametrization of this form factor
(the pole variant) discussed in \cite{UKQCD}. 
The quark model estimate makes use of the single-pole
approximation $f^0(p^2) = f^+(0)/(1 - m_0^2/p^2)$ with $m_0 = 6.0$ GeV.
In \cite{DKS} it is suggested to use instead the relation (\ref{CT}) 
in order to normalize $f^0$  at maximum momentum transfer. 
Extrapolation to intermediate and small 
$p^2$ then leads to a result very similar to the one obtained in 
\cite{BSW} and shown in the figure.
Despite of the rough agreement with our sum rule result we doubt the 
validity of the single-pole model at small $p^2$ for reasons explained
in sect. 5. Finally, in the framework of the heavy-quark-effective-theory
one has derived a three-point sum rule for $f^0$ \cite{ColSan}
giving a result in the region  $0 < p^2 < 10$ GeV$^2$ which is
about  $30\%$ lower than the expectation from the light-cone sum rule.
Concerning similar applications of the light-cone sum rules, 
one should also mention the study of the form factor $f^-$ of 
the $B \to K$ transition in \cite{aliev}. 

It would be very interesting to confront these predictions with 
experimental data, not only to test the theoretical methods
of calculating form factors, but also since $f^0$ enters
the factorized amplitudes for a class of nonleptonic two-body
decays. Yet, direct measurements of $f^0$ are only feasible 
in the semileptonic decay $B^0\rightarrow\pi^- \tau^+ \nu_\tau$.
This mode may get in experimental reach at future $B$ factories. 
We have presented the expected decay spectra and 
demonstrated the sensitivity to $f^0$.     

Last but not least, light-cone sum rules for heavy-to-light
form factors such as $f^0$ provide very flexible tools
to study the transition from small to large momentum transfers,
and to relate the physics of $D$ and $B$ mesons. Moreover,
they provide new insights in the heavy quark mass dependence of
weak matrix elements.

\section{Acknowledgements}

We are grateful to 
V.M. Braun and O. Yakovlev for very useful discussions. 
This work was supported by the Bundesministerium f\"ur
Bildung, Wissenschaft, Forschung und Technologie, Bonn, Germany,
Contract 05 7WZ91P(0).

\section*{Appendix}
\app
Here, we give the explicit expressions 
for the surface term $t^+$ and 
the contribution $f_G^+$ from the quark-antiquark-gluon wave functions 
in the light-cone sum rule (\ref{fplus}):
$$
t^+(s_0^B,p^2,M^2)=
\exp\left(-\frac{s_0^B}{M^2}\right)
\Bigg\{
\frac{\mu_\pi(m_b^2+p^2)}{6m_b(m_b^2-p^2)}
\varphi_\sigma(\Delta)
$$
$$
-\frac{4m_b^2}{(m_b^2-p^2)^2}\left(
1+\frac{s_0^B-p^2}{M^2}\right) g_1(\Delta)
+
\frac{4m_b^2 }{(s_0^B-p^2)(m_b^2-p^2)}
\frac{dg_1(\Delta)}{du}
$$
\be
+\frac2{m_b^2-p^2}\Bigg[1+\frac{m_b^2+p^2}{m_b^2-p^2}\left(
1+\frac{s_0^B-p^2}{M^2}\right) 
\Bigg]\int^\Delta_0g_2(v)dv-
\frac{2(m_b^2+p^2)}{(m_b^2-p^2)(s_0^B-p^2)}g_2(\Delta)\Bigg\}
\label{t+}~,
\ee

$$
f_G^+(p^2,M^2)=
-\int_0^1\!\!u du\!\int \frac{{\cal D}\alpha_i
\Theta( \alpha_1+u\alpha_3-\Delta)}{(\alpha_1+u\alpha_3)^2}
\exp\!\left(-\frac{m_b^2-p^2
(1-\alpha_1-u\alpha_3)}{(\alpha_1+
u\alpha_3)M^2}\right)\!
$$
$$
\times
\Bigg\{\frac{2f_{3\pi}}{f_{\pi}m_b} 
\varphi_{3\pi}(\alpha_i)
\left[1-\frac{ m^2_b -p^2 }{(\alpha_1+u\alpha_3)M^2}\right]    
\nonumber
$$
\be
-\frac1{uM^2} \Bigg[2\varphi_\perp (\alpha_i)-\varphi_\parallel (\alpha_i)+
2\tilde{\varphi}_\perp (\alpha_i)-\tilde{\varphi}_\parallel (\alpha_i)\Bigg]
\Bigg\}~,
\label{formSR}
\ee
with ${\cal D}\alpha_i= d\alpha_1d\alpha_2d
\alpha_3\delta(1-\alpha_1-\alpha_2-\alpha_3)$.
The definitions and functional forms of the twist 3 wave function 
$\varphi_{3\pi}$ and the twist 4 wave functions $\varphi_\perp$
$\varphi_\parallel$, $\tilde{\varphi}_\perp$ and $\tilde{\varphi}_\parallel$
can be found in \cite{BBKR,BF}. 

\appende

\pagebreak                     


\begin{figure}[htb]
\centerline{
\epsfig{bbllx=100pt,bblly=209pt,bburx=507pt,%
bbury=490pt,file=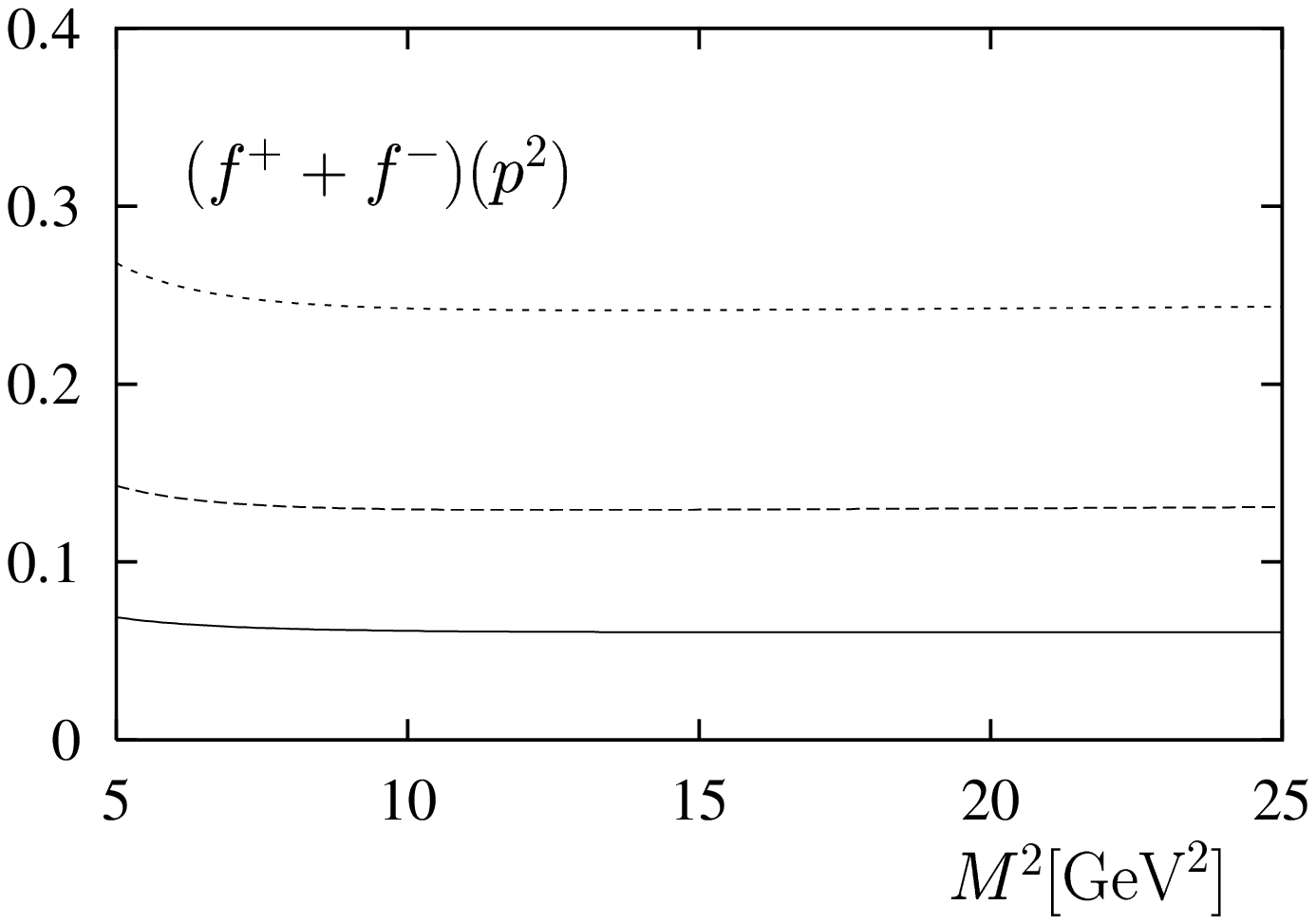,scale=0.9,%
clip=}
}
\caption{\it 
Form factor  $(f^++f^-)$ as a function of the 
Borel parameter at various values of the momentum transfer: 
$p^2=0$ (solid), $p^2=10$ GeV$^2$ (long-dashed) and $p^2=16$ GeV$^2$ (short-dashed).}
\end{figure}

\begin{figure}[htb]
\centerline{
\epsfig{bbllx=100pt,bblly=209pt,bburx=507pt,%
bbury=490pt,file=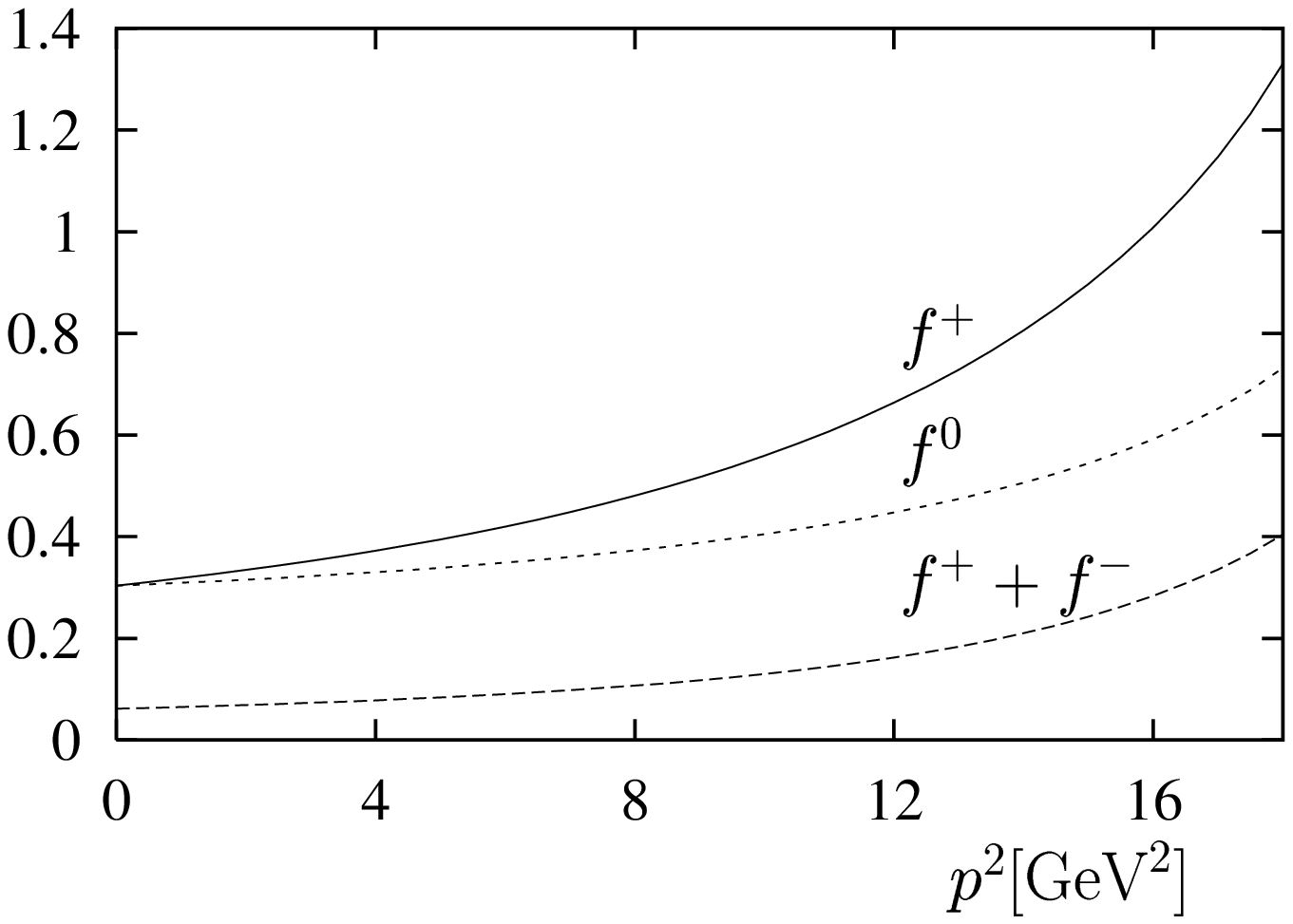,scale=0.9,%
clip=}
}
\caption{\it $B \to \pi$ form factors
obtained from light-cone sum rules.}
\end{figure}

\begin{figure}[htb]
\centerline{
\epsfig{bbllx=100pt,bblly=209pt,bburx=507pt,%
bbury=490pt,file=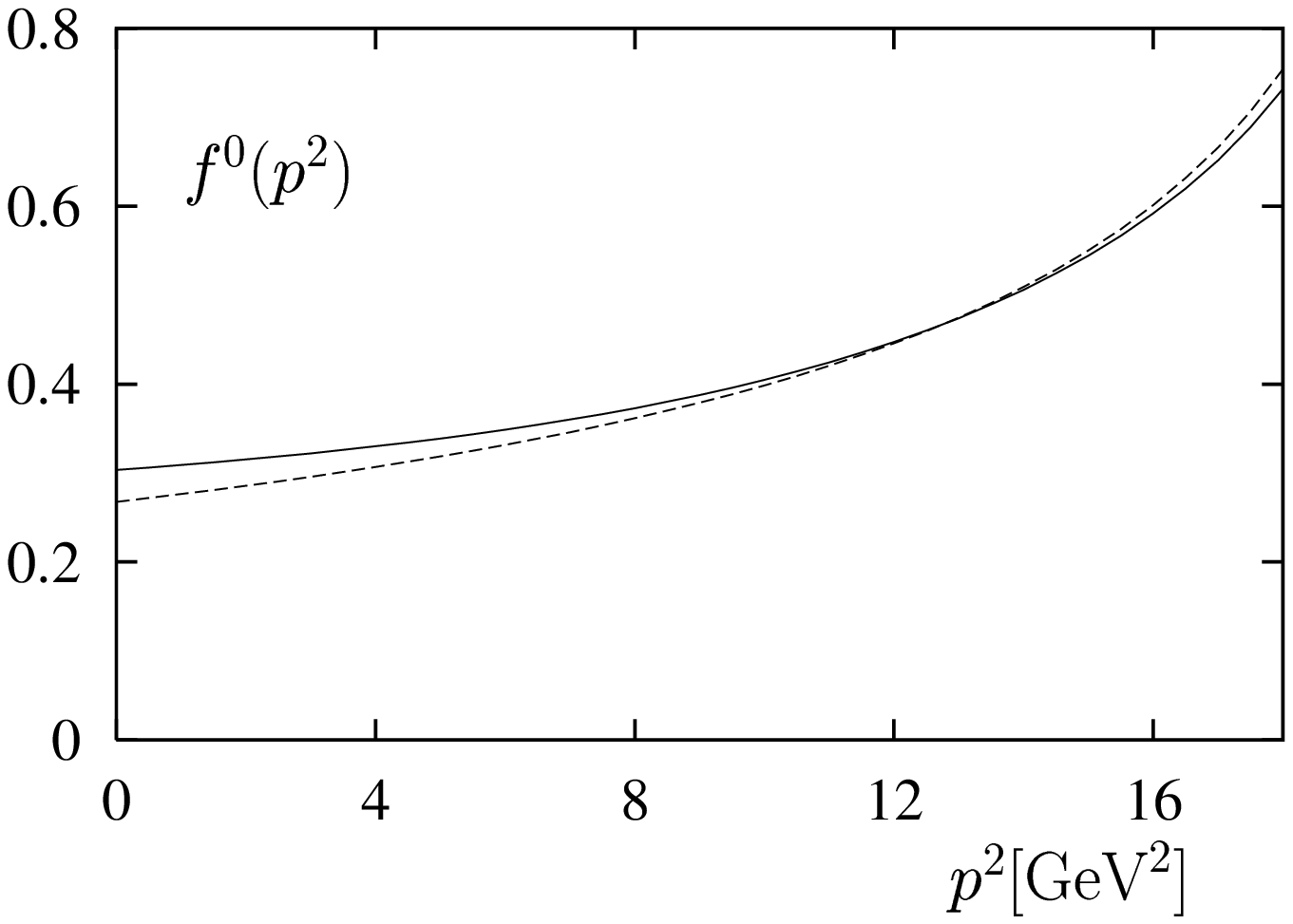,scale=0.9,%
clip=}
}
\caption{\it 
Sensitivity of the form factor $f^0$ to the 
light-cone wave functions: nonasymptotic corrections 
included (solid) and purely asymptotic w.f. (dashed)} 
\end{figure}

\begin{figure}[htb]
\centerline{
\epsfig{bbllx=100pt,bblly=209pt,bburx=507pt,%
bbury=490pt,file=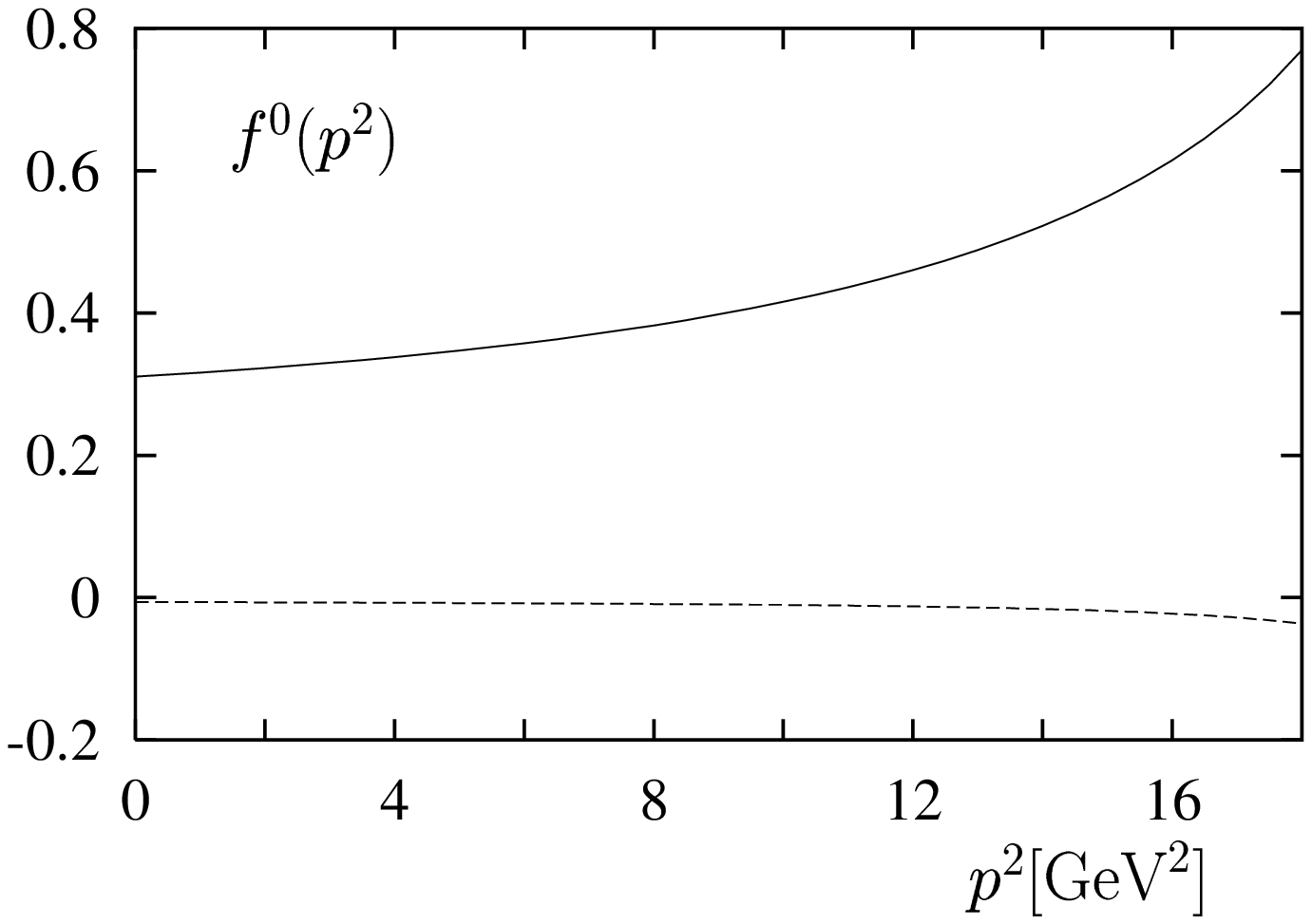,scale=0.9,%
clip=}
}
\caption{\it $B \to \pi$ form factor $f^0$: twist 2 and 3 
contributions  (solid), twist 4 contribution (dashed).
}
\end{figure}

\begin{figure}[htb]
\centerline{
\epsfig{bbllx=100pt,bblly=209pt,bburx=507pt,%
bbury=490pt,file=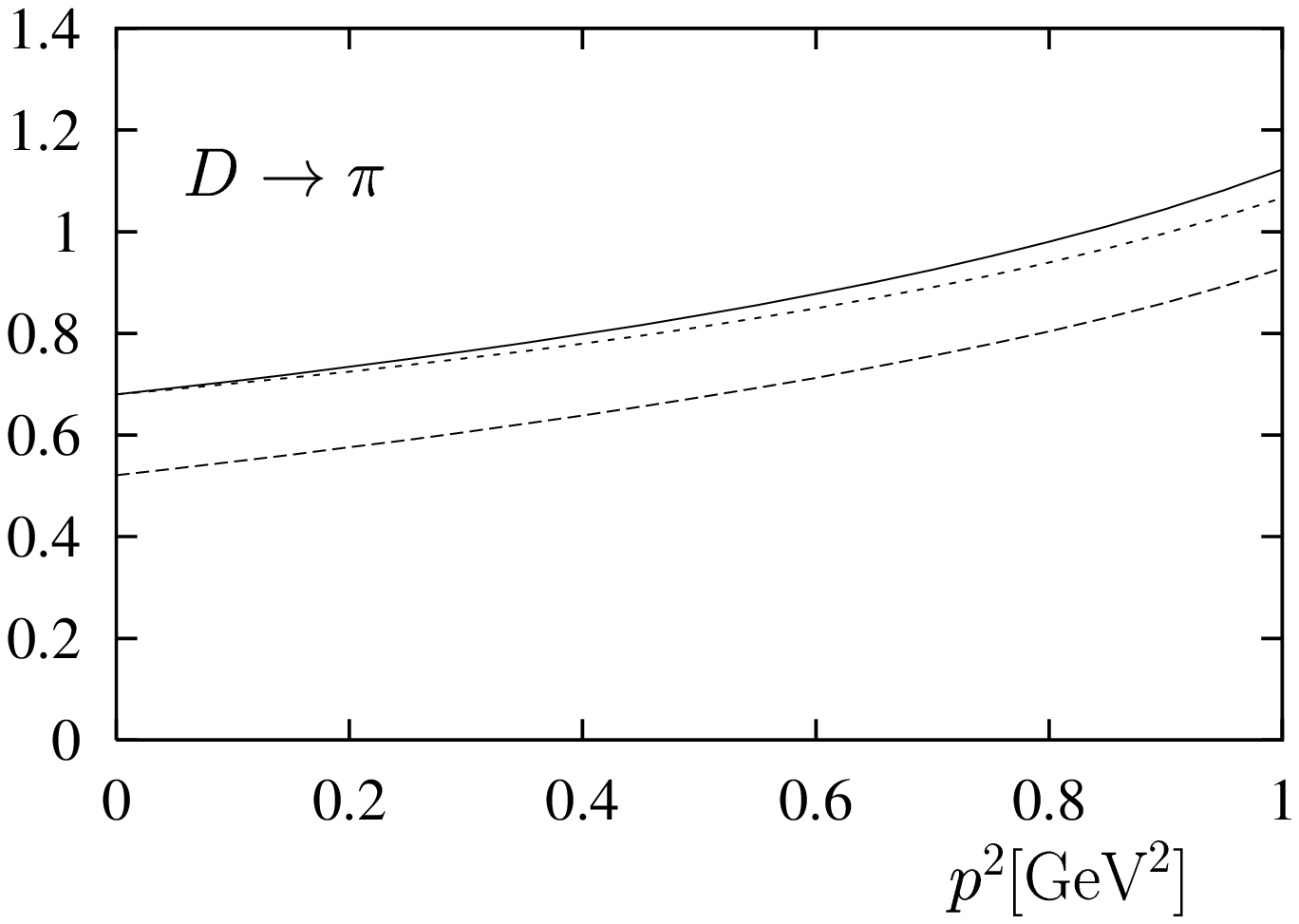,scale=0.9,%
clip=}
}
\caption{\it $D \to \pi$ form factors
obtained from light-cone sum rules: $f^+$ (solid), $f^+ + f^-$ (dashed)
and $f^0$ (dotted).}
\end{figure}

\begin{figure}[htb]
\centerline{
\epsfig{bbllx=100pt,bblly=209pt,bburx=507pt,%
bbury=490pt,file=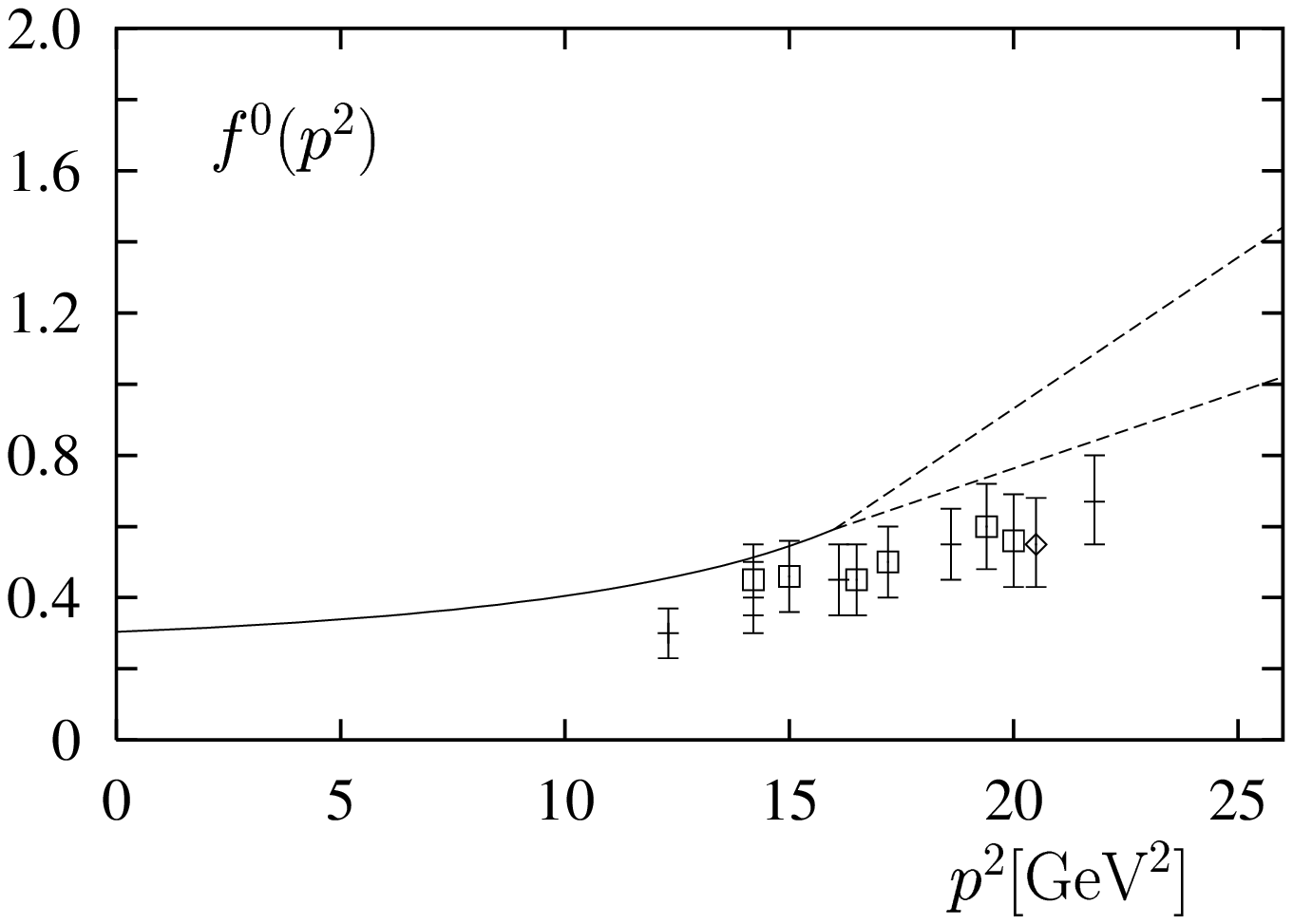,scale=0.9,%
clip=}
}
\caption{\it 
The $B\to \pi$ form factor $f^0$ : direct sum rule estimate (solid) 
and linear extrapolations to the limit (\ref{CT}) (dashed).
The lattice results are from \cite{Flynn}.}
\end{figure}

\begin{figure}[htb]
\centerline{
\epsfig{bbllx=100pt,bblly=209pt,bburx=507pt,%
bbury=490pt,file=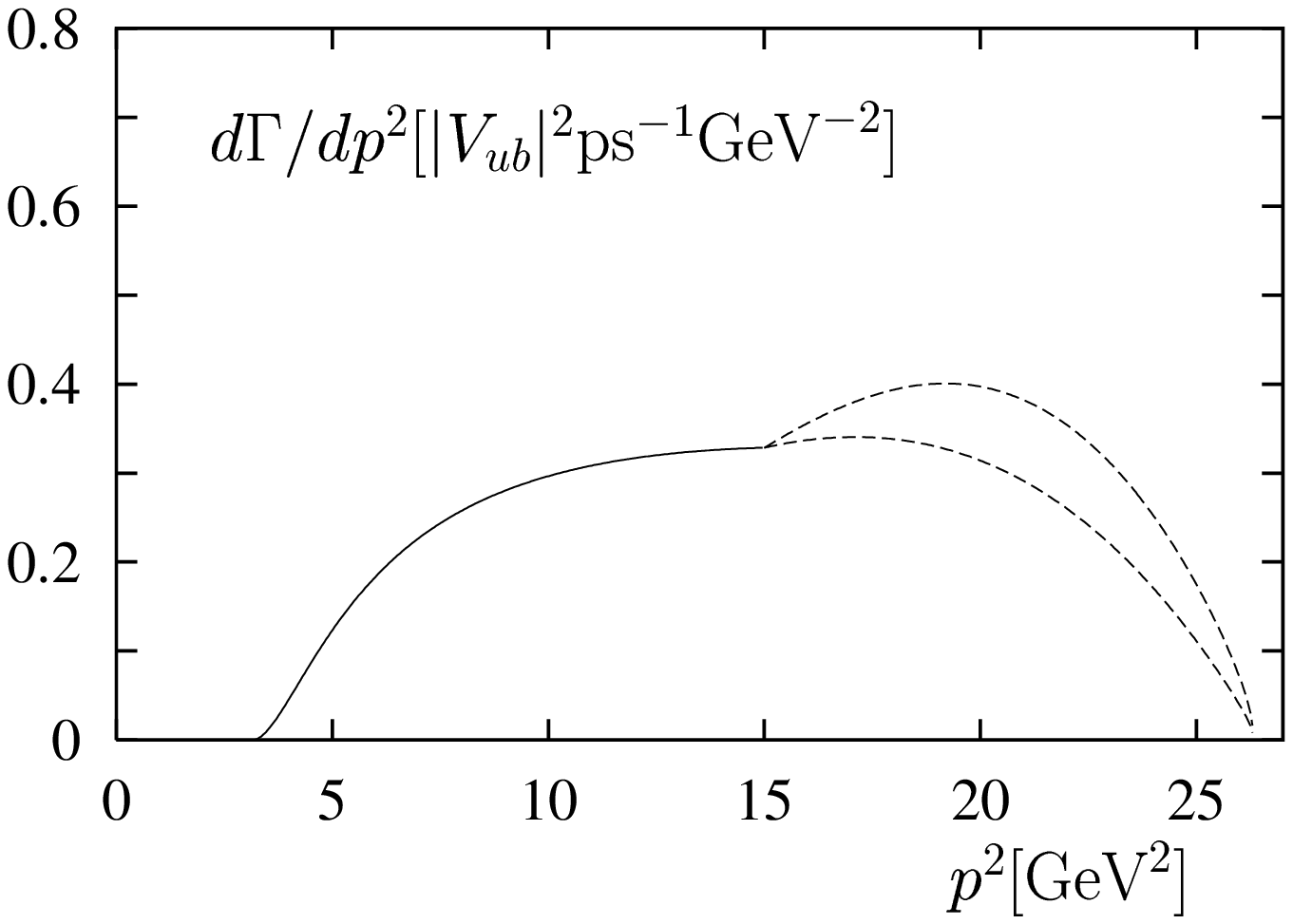,scale=0.9,%
clip=}
}
\caption{\it 
Distribution of the momentum transfer squared in 
 $B\to \pi \bar{\tau}\nu_\tau$. The two curves 
correspond to the two extrapolations of $f^0$ shown in 
Fig. 6.}
\end{figure}

\begin{figure}[htb]
\centerline{
\epsfig{bbllx=100pt,bblly=209pt,bburx=507pt,%
bbury=490pt,file=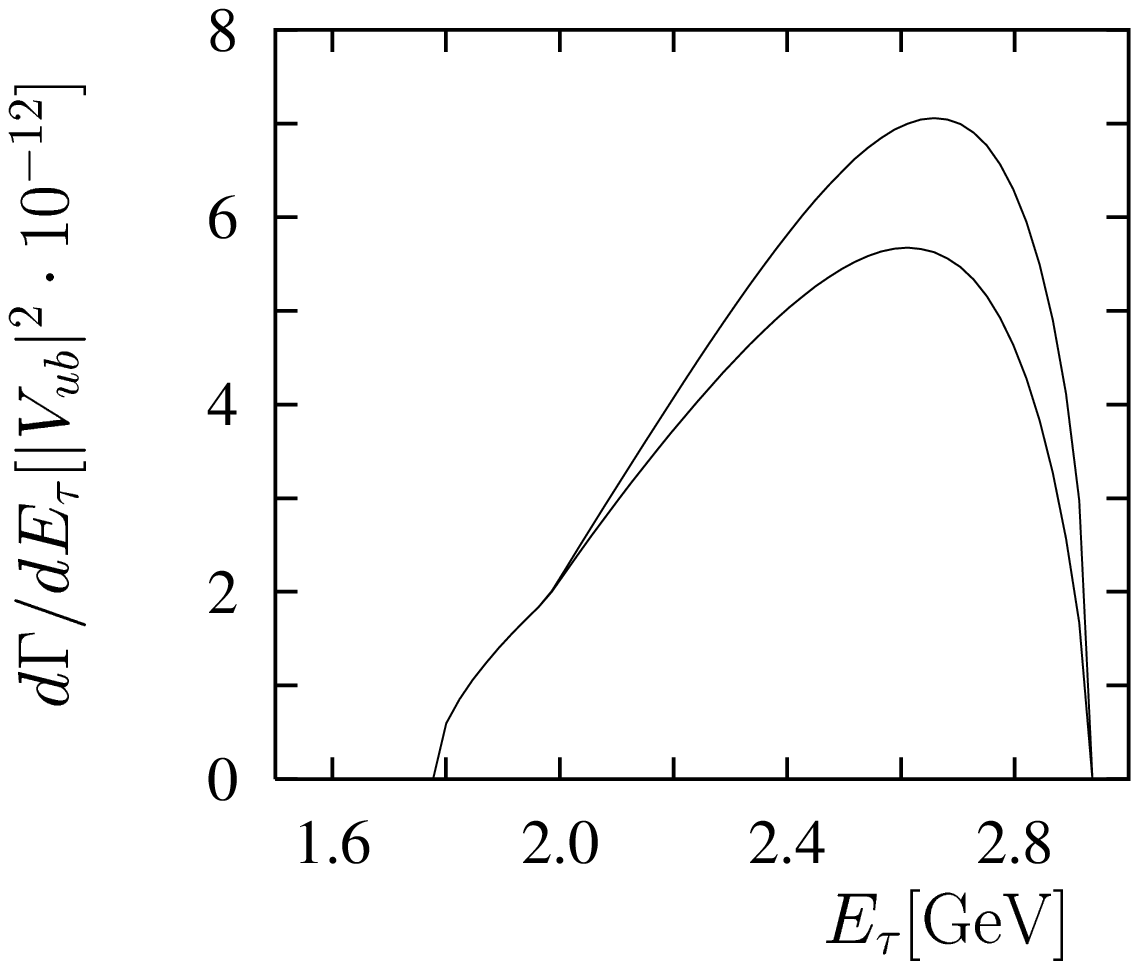,scale=0.9,%
clip=}
}
\caption{\it 
Distribution of the $\tau$-lepton energy in $B\to \pi \bar{\tau}\nu_\tau$.
The two curves correspond to the two extrapolations of $f^0$ shown in 
Fig. 6.}
\end{figure}

\begin{figure}[htb]
\centerline{
\epsfig{bbllx=100pt,bblly=209pt,bburx=507pt,%
bbury=490pt,file=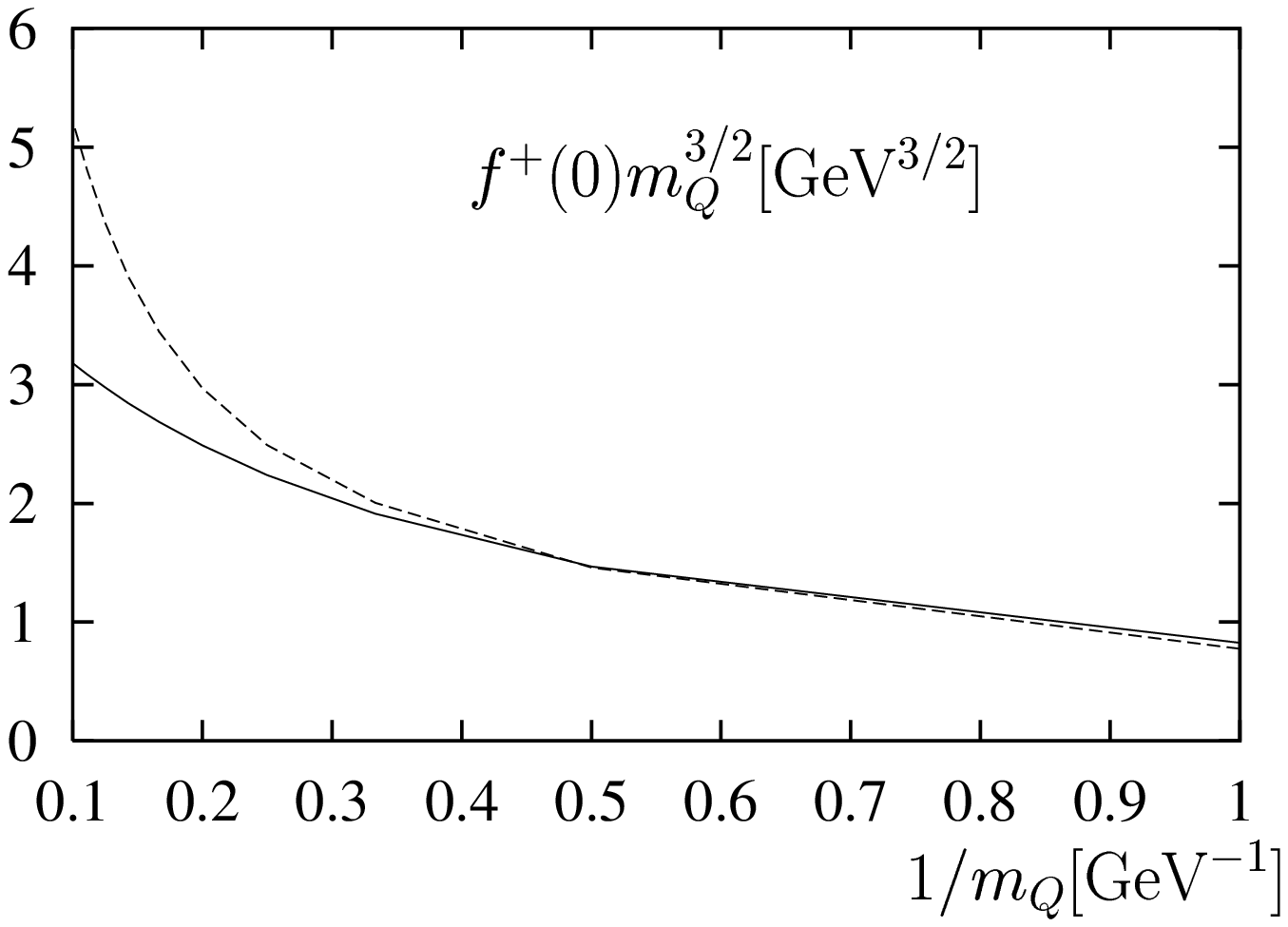,scale=0.9,%
clip=}
}
\caption{\it 
Dependence of the form factor $f^+(0)$ on the heavy quark mass $m_Q$ for 
purely asymptotic wave functions (solid), and the nonasymptotic corrections
included at the scale $\mu_Q =\sqrt{2m_Q\bar{\Lambda}}$  
(dashed).}
\end{figure}

\begin{figure}[htb]
\centerline{
\epsfig{bbllx=100pt,bblly=209pt,bburx=507pt,%
bbury=490pt,file=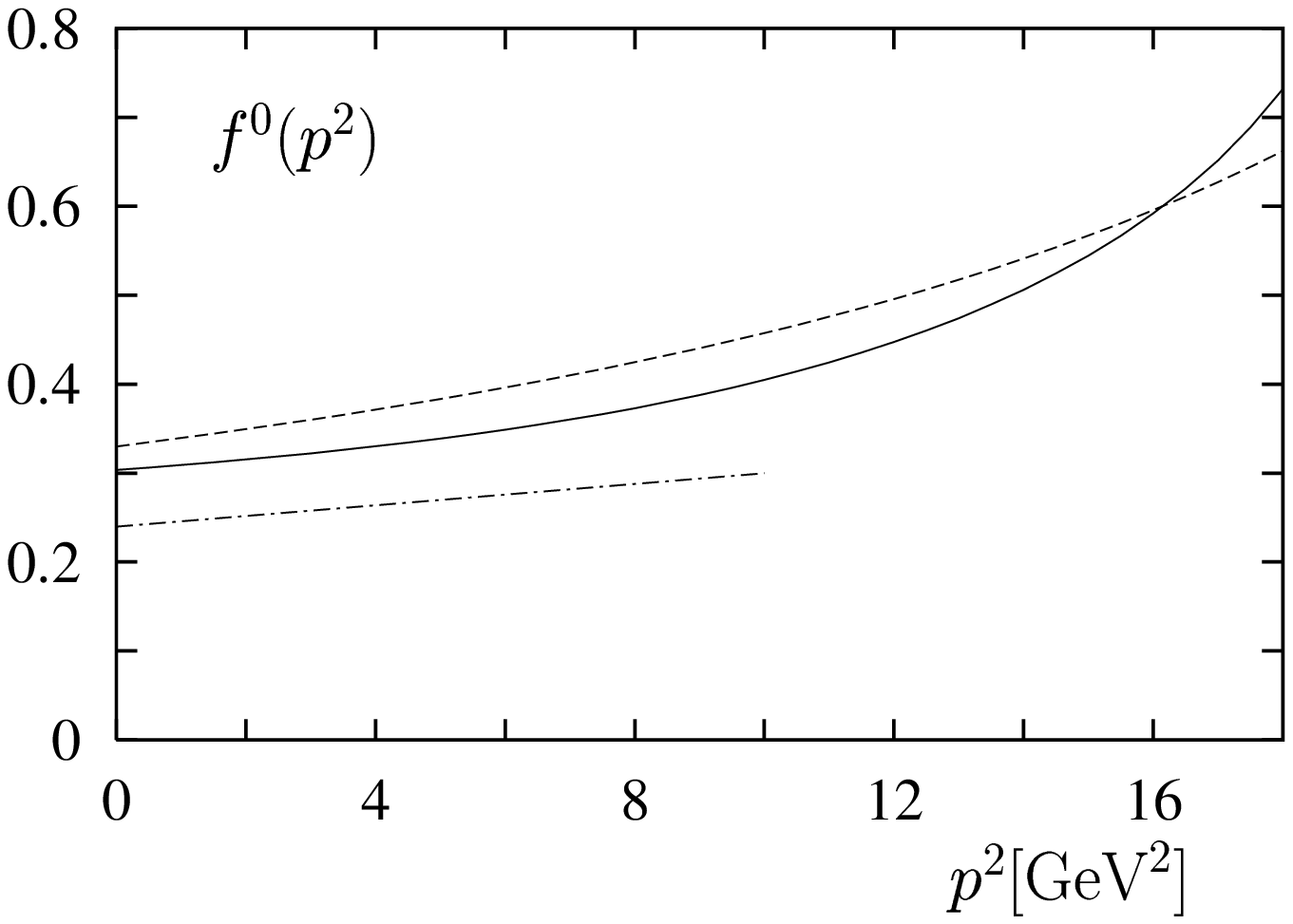,scale=0.9,%
clip=}
}
\caption{\it 
The $B \rightarrow \pi$   form factor $f^0$:  
light-cone sum rule (solid) in comparison to  
the quark model prediction from
\cite{BSW}  (dashed)  and the QCD sum rule result from
\cite{ColSan} (dash-dotted).}

\end{figure}

\end{document}